\documentclass[pra, twocolumn,aps,superscriptaddress,showpacs]{revtex4-2} 
\usepackage{tabularx} 
\usepackage{amsmath}  
\usepackage{graphicx} 
\usepackage[margin=1in,letterpaper]{geometry} 
\usepackage{natbib}
\usepackage[final]{hyperref} 

\begin{document}

\title{Improving Sensitivity of an Amplitude-Modulated Magneto-Optical Atomic Magnetometer using Squeezed Light}
\author{Jiahui Li}
\author{Irina Novikova}
\address{Department of Physics, College of William \& Mary, Williamsburg, VA 23187 USA}

\begin{abstract}
We experimentally demonstrate that a squeezed probe optical field can improve the sensitivity of the magnetic field measurements based on nonlinear magneto-optical rotation (NMOR) with an amplitude-modulated pump when compared to a coherent probe field under identical conditions. To realize an all-atomic magnetometer prototype, we utilize a nonlinear atomic interaction, known as polarization self-rotation(PSR), to produce a squeezed probe field. An independent pump field, amplitude-modulated at the Larmor frequency of the bias magnetic field, allows us to extend the range of most sensitive NMOR measurements to sub-Gauss magnetic fields. While the overall sensitivity of the magnetometer is rather low ($>250~\mathrm{pT}/\sqrt{\mathrm{Hz}}$, we clearly observe a $15~\%$ sensitivity improvement when the squeezed probe is used. Our observations confirm the recently reported quantum enhancement in a modulated atomic magnetometer~\cite{TroullinouPRL2021}.  
\end{abstract}
\maketitle

Optical atomic magnetometers~\cite{Budker2013,Grosz2017,Weis2017} – devices that enable magnetic field measurements using changes in optical properties of an atomic medium – are becoming one of the most promising platforms for precise magnetic field measurements, and are finding wide applications including geophysics~\cite{Dang2010,Zhang2021PRL}, medical diagnosis~\cite{Bison2003,Xia2006,Lembke2014,Borna2020}, fundamental science~\cite{JacksonKimball2017,Afach2018,Rosner2022}, and many others~\cite{atomicmagnJNPreview2022}. The general operational principle for such devices is based on the strong relationship between optical absorption and refraction and the spin orientation or alignment of ground-state electrons in atoms. Once the atomic spins are initialized by an optical field, their evolution in the magnetic field is monitored via optical means, giving direct access to the magnetic field measurements. One of the common diagnostic approaches relies on the magnetic-optical polarization rotation (NMOR), in which the magnitude of the magnetic field is calculated from the measured rotation angle of the linear polarization of an optical probe field~\cite{Budker2002RMP}.  This detection method is utilized in some of the most sensitive atomic magnetometers~\cite{Dang2010,PattonPRL2014,Li2018}. 

\begin{figure}[ht!]
    \centering
    \includegraphics[width = 0.8\columnwidth]{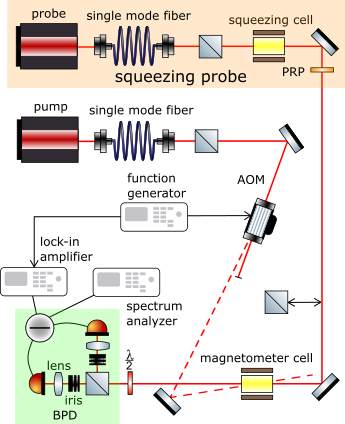}
    \caption{The schematic of the experimental setup. Two independent lasers are used to produce the pump and probe fields. Polarization squeezing for the probe optical field is generated in an auxiliary ${}^{87}$Rb (PSR) cell before entering the main magnetometer cell, and its quadrature is controlled with a phase retardation plate (PRP). The pump field is amplitude modulated using an acousto-optical modulator (AOM). The dynamic changes in the probe field polarization are detected using a balanced photodetector (BPD), and analyzed either using a spectrum analyzer, or a lock-in amplifier. }
    \label{fig:Schematics}
\end{figure}

Once all technical noises are suppressed, the ultimate sensitivity of the polarization measurements is limited by the projection noise of an atomic ensemble~\cite{Sewell2012,Rochester2012} and by the shot noise of an optical probe~\cite{Lucivero2014}. To improve the performance beyond these limits, quantum mechanical tools become necessary. In the regime where the magnetometer performance is shot-noise limited, the use of squeezed optical probe becomes beneficial. Previous experiments utilized a linearly polarized optical probe field in which the coherent vacuum in the orthogonal polarization was replaced by the squeezed vacuum. They demonstrated the improved magnetometer sensitivity for ac magnetic field in the spectral range in which squeezing was observable~\cite{Wolfgramm2010,Horrom2012,OtterstromOL2014,BaiJOpt2021}. Technical $1/f$ noise has prevented detection of slowly varying dc magnetic fields. 

Recently, the quantum-enhanced measurements of $43~\mu$T magnetic field with sub-pT precision have been demonstrated ~\cite{TroullinouPRL2021}. In this work, the problem of the $1/f$ noises was solved by using an additional pump field modulated at Larmor frequency. Such modulation shifts the dc-magnetic response into a high-frequency domain~\cite{Budker2002PRA,WeisPRA2013}, where the technical noises are reduced, and the magnetometer performance is mostly limited by the fundamental quantum noises. That experiment demonstrated 20~\% improvement in magnetometer sensitivity, when shot-noise limited, using a crystal-based source of polarization squeezed optical probe. Additional demonstrated advantages of this approach were the increased measurement bandwidth and elimination of the back-action noise at higher atomic densities.

Here we investigate the possibility of realization of an all-atomic quantum enhanced magnetometer by using Rb atomic vapor cells for both magnetic field measurements and to produce a non-classical optical probe. For the latter, we exploit a source of squeezed vacuum based on polarization self-rotation mechanism (so-called PSR-squeezed vacuum~\cite{matskoPRA2002,mikhailov2008ol}). We show that the use of a 2dB on-resonance squeezed probe increases the signal to noise (SNR) ratio of the magnetic response at the modulation frequency, thus increasing the estimated magnetometer sensitivity by 15\%. This result confirms the quantum enhanced modulated magneto-optical magnetometry reported in \cite{TroullinouPRL2021}, while operating in a very different parameter subspace.  In addition to a different source of squeezed light, we also employed an amplitude-modulated, rather than phase-modulated, pump field, as shown in Fig.~\ref{fig:Schematics}, inspired by the AMOR magnetometer configuration~\cite{Gawlik2006,WeisPRA2013,Lucivero2014,Li2020}. 
We also aimed to achieve a high operational dynamic range by using broad optical resonances capable of operating in terrestrial magnetic fields~\cite{Li2020}. For this purpose, we operated at a rather high probe field power. Also, since PRS squeezing exists only at certain optical transitions in Rb, we relied on high probe power and optical saturation to minimize resonant optical losses. We also used a vapor cell with a small amount of buffer gas (and potentially without buffer gas), as velocity-changing collisions at high buffer gas pressure cells are known to reduce the magneto-optical rotation~\cite{Novikova2005}. Unsurprisingly, we found that in this regime the overall magnetic sensitivity was rather low due to the short transient spin coherence lifetime. Nevertheless, we observed a clear improvement with the use of the squeezed probe field. Our results may be of interest for applications that cannot use buffer gas to restrict Rb motion or need to operate outside magnetic shielding. Also, the simplicity of the experimental apparatus offers some practical appeal. 


In this experiment, we employ two independent external cavity diode lasers. The probe beam was tuned near the $5S_{1/2}F = 2 \rightarrow 5P_{1/2}F' = 2$ transition to maximize the amount of PSR squeezing. A detailed description of the PSR squeezing apparatus is provided in Ref.\cite{mikhailov2008ol,MiZhangPRA2016}, but in short, the propagation of a linearly polarized light through a $^{87}$Rb vapor cell produced $\approx 2$~dB of a quadrature-squeezed vacuum in the polarization orthogonal to the original laser field. For the further measurements, we treated the total output as a polarization-squeezed optical field~\cite{BarreiroPRA2011}. The phase retardation plate PRP (a wave plate tilted to add a controllable relative phase between the two polarization components), placed after the squeezing cell, was used to control the squeezing quadrature and adjust it to minimize the fluctuations in the polarization rotation angle. Achieving the maximum squeezing level required operation at relatively high laser power ($6.5$~mW). We did not take any additional steps to reduce the probe beam power before the magnetometery cell. However, it is straightforward to reduce it without compromising the amount of noise suppression using a simple Mach-Zender interferometer~\cite{CuozzoAQT2022}, and our measurements suggest that the magnetometer sensitivity may be improved with additional optimization of the mean probe power.
 
A separate external cavity diode laser produced the pump field, and we used an acousto-optical modulator (AOM) with 50\% duty cycle to modulate its power at a frequency close to the Larmor frequency of the longitudinal bias magnetic field $\Omega_L = \gamma B_z$, where $\gamma$ is the gyromagnetic ratio. Such modulation creates stroboscopic optical pumping of Rb atoms, that in turn creates a strong magneto-optical polarization response at the modulation frequency in the (unmodulated) probe optical field~\cite{Gawlik2006,WeisPRA2013,Lucivero2014,Li2020}. We varied the pump power from 0.6~mW to 5.6~mW, but since a stronger pump field amplified the desired magneto-optical response with little impact on the noise level, we conducted the measurements using the largest available power of 5.6~mW. 
We also optimized the detuning of the pump beam: while we observed the modulation-induced polarization rotation resonances for the pump field tuned to every optical resonance of the $D_1$ Rb line, we found that the magnetometer response is significantly stronger when it was tuned to the $5S_{1/2}F = 1 \rightarrow 5^2 P_{1/2} F' = 2$ transition. 

To suppress possible leakage of the pump field into the probe detection, the beams are counter-propagating when crossing inside the main magnetometer cell at an angle of $13$ degrees. This geometry opens an interesting possibility of measuring a local value of the magnetic field corresponding only to the region in which the two optical fields intersect, thus enabling a spatial mapping of the magnetic field.

\begin{figure}[ht!]
    \centering
    \includegraphics[width = 1.0\columnwidth]{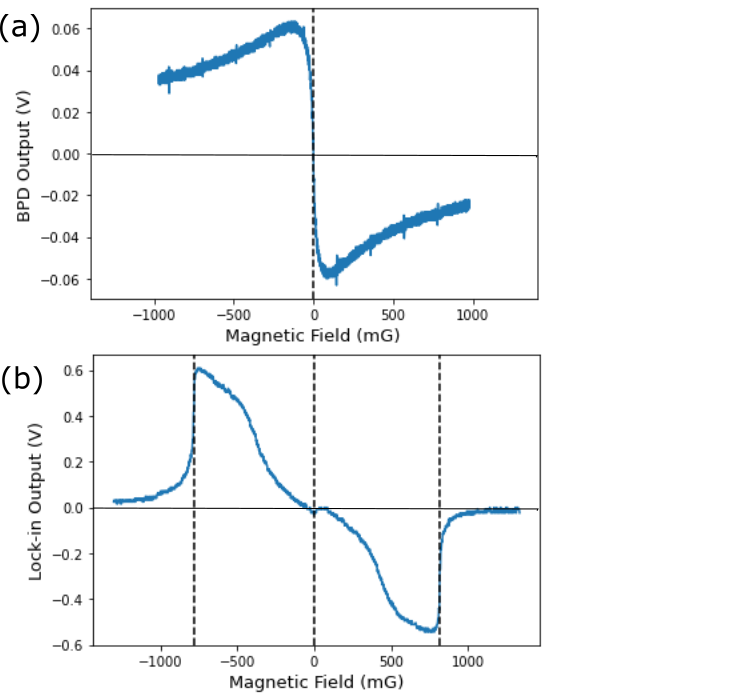}
            \caption
        {(a) Direct differential photodetector output as a function of an applied longitudinal magnetic field. The recorded PBD output is proportional to the polarization rotation angle. (b) Corresponding lock-in amplifier output versus the magnetic field. The modulation frequency is set to be 580 kHz corresponding to 800 mG magnetic field. The lock-in phase is adjusted to maximize the side resonances.}
       \label{fig:originalsignal}
\end{figure}

For the magnetic field measurements, we used a Pyrex vapor cell (length 75~mm, 25~mm diameter) containing isotopically enriched $^{87}$Rb and 2.5~Torr of Ne buffer gas. The cell was mounted inside a magnetic shielding while the longitudinal magnetic field $B_z$ was created by an internal solenoid. Unless otherwise specified, the cell temperature was maintained at 40.3$^\circ$C corresponding to an atomic density of $5.5\cdot 10^{10}~\mathrm{cm}^{-3}$. 
The changes in the probe polarization rotation were monitored using a balanced polarization detector (BPD), consisting of a half-wave plate, orienting the probe polarization at 45$^\circ$ to the axis of a polarization beamsplitter, and the two identical photodiodes monitoring light power in each of the beamsplitter output channels. The photodetector photocurrents were subtracted, so that the output BPD voltage was proportional to the polarization rotation angle. An example of polarization rotation spectra of the probe field without the modulated pump is shown in Fig.~\ref{fig:originalsignal}(a). As expected, the polarization rotation angle as a function of the applied magnetic field presents a dispersion-like curve, and changes most rapidly around zero magnetic field within a few tens of mG. 

The interaction of atoms with the modulated pump field creates a component of the atomic coherence oscillating at the pump modulation frequency $\Omega_m$, which consequently creates an optical response at that frequency in the unmodulated probe field. What is critical for the magnetometer operation is that the magnitude of this response increases dramatically when the modulation frequency matches the Larmor precession frequency  $\Omega_m \approx\Omega_L$, so that the time-dependent atomic coherence variation induced by the pump field modulation and by the Larmor precession constructively interfere.  
In this case, in addition to the stationary response, we observed a periodic variation of the polarization rotation at $\Omega_m$. This high-frequency response can be easily isolated using a lock-in amplifier. Fig.~\ref{fig:originalsignal}(b) shows the example of lock-in output for the pump modulation frequency set to $580$~kHz and slowly varying magnetic field. One can clearly see the appearance of two sharp resonances at the magnetic field values about $B_z=\pm 800$~mG, corresponding to the $580$~kHz Larmor frequency. The lock-in phase is adjusted to maximize the slope of the modulation-induced resonances near $\Omega_m\approx \Omega_L$. This high response allows us to detect small changes in the magnetic field on the top of the large mean value. 


\begin{figure}
          \centering
        \includegraphics[width=0.9\columnwidth]{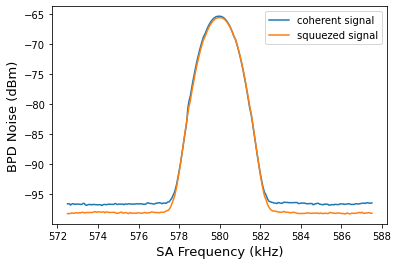} 
        \caption
        {Spectrum analyzer signal for the BPD output near the pump modulation frequency $\Omega_m  = 580~\mathrm{kHz} \approx \Omega_L$ for the coherent and squeezed probe field. The spectrum analyzer resolution and video bandwidth values are 1000~Hz and 1~Hz respectively. }
        \label{fig:SA}
\end{figure}

In the parameter space of this experiment, the leading source of noise in the polarization rotation measurements was the photon shot noise, making the use of the squeezed light beneficial for reducing the noise level and potentially improving the magnetometer sensitivity. To confirm this experimentally, we compared the noise level of the magnetic field measurements performed with squeezed or coherent probe beams.
To switch between these two cases we used a polarizing beam splitter aligned to reject the polarization of the PSR squeezed vacuum: when inserted into the probe beam, it effectively ``converted'' the squeezed optical field into a coherent one without affecting its other parameters. 
One of the features of PSR squeezing is that it occurs only near the center of an atomic resonance~\cite{mikhailov2008ol,BarreiroPRA2011,Horrom2012}. On one hand, it automatically ensures the strongest coupling to the atomic transition; on the other hand, it unavoidably raises the issue of resonant absorption and its negative effect on squeezing. In this experiment, this absorption was greatly reduced by optical saturation thanks to a relatively high probe power. We observed about  10\% absorption by the magnetometer Rb cell (including some residual reflection on cell windows), resulting in squeezing attenuation of 0.3~dB, from -1.9~dB to -1.6~dB squeezing before and after the cell. 

Fig.~\ref{fig:SA} shows the sample spectrum analyzer traces of the BPD output for squeezed and coherent probe fields. The central peak at the pump field modulation frequency $\Omega_m=580$~kHz represents the magnitude of the induced probe polarization rotation. The height of this peak is proportional to the power of the probe optical field, as well as to the difference between $\Omega_L$ and $\Omega_m$, since the atomic response to the modulation is the strongest near the resonance condition, as discussed above. The noise floor in our case was determined by the fluctuations of the probe field. For a coherent probe field, we confirmed that the measurements were shot-noise limited. If a squeezed probe was introduced, the measured noise was determined by the noise quadrature measured by the BPD. We controlled this by adjusting the phase retardation plate, for all the presented measurements the quantum fluctuations of the detected probe polarization rotation were minimized. Fig.~\ref{fig:SA} shows the $1.6$~dB drop in the noise level between the coherent and squeezed probe, as anticipated. 

\begin{figure}
    \centering
   \includegraphics[width = 1.0\columnwidth]{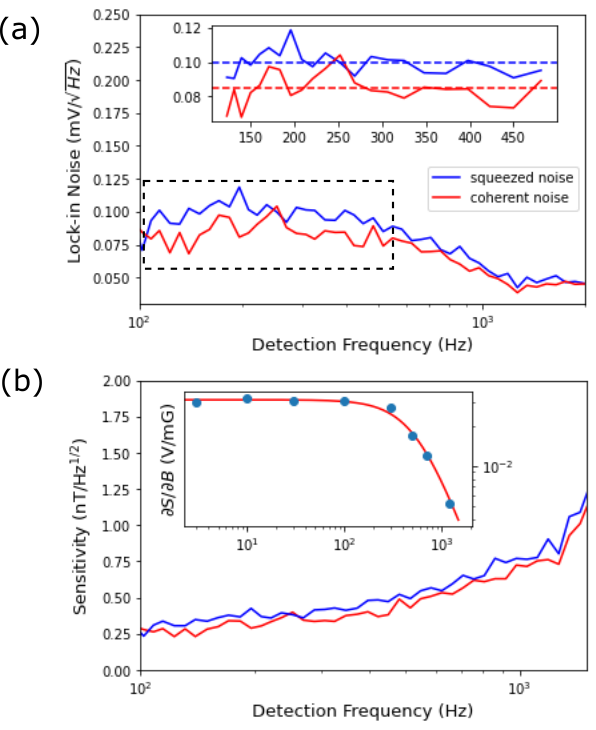}
    \caption{(a) The noise spectra of the lock-in output at the magnetometer operation point ($\Omega_m=\Omega_L$) for coherent and squeezed probe fields (squeezing levels are -1.9~dB/-1.6~dB before/after the magnetometer cell). The inset shows the zoomed-in section with average noise levels for the two curves. (b) The estimated sensitivity of the magnetic field measurements for coherent and squeezed probe fields. The inset shows the measured response spectrum $\partial S/ \partial B (f)$ of the magnetometer as a function of the detection frequency $f$. The higher-frequency roll-off is due to the limited minimum $0.3$~ms integration constant of the lock-in amplifier.  }
    \label{fig:sensitivity}
\end{figure}

The maximum magnetometer sensitivity is achieved for the phase-sensitive detection with the optimized detection phase, so it was important to verify that the noise reduction near that modulation frequency translates into reduced noise of the lock-in readout. In this case, we estimated the magnetic response $\partial S/ \partial B $ by measuring the slope of the lock-in discrimination curve near $\Omega_L$. The noise spectrum of the lock-in data $\sigma_S(f)$  is shown in Fig.~\ref{fig:sensitivity}(a). At the low frequencies ($<100Hz$) the noise was dominated by the fluctuations of the magnetic field and other technical noises. The limited minimum integration constant of the lock-in amplifier ($300~\mu$s) caused the slow roll-off at the higher frequencies. However, in the intermediate region, we observe a clear reduction of the measured noise with the squeezed probe compare to the coherent field. 

Using these measured values, we estimated the sensitivity of such a magnetometer $\delta B_{min}(f)$ for various detection frequencies $f$ as:  
\begin{equation}
\delta B_{min}(f) = \frac{\sigma_S(f)}{\partial S/ \partial B}. 
\end{equation}
To account for the reduction of the signal at higher frequencies we calibrated its frequency response by adding a small modulation to the applied magnetic field, and recording its amplitude in the lock-in readout. The resulting dependence is shown as an inset in Fig.~\ref{fig:sensitivity}(b) where we fit it to the Eq.(3) of Ref.~\cite{TroullinouPRL2021}. As expected, the overall sensitivity is rather low (around $250~\mathrm{pT}/\sqrt{Hz}$ due to the large width of the magneto-optical resonance. However, one can clearly observe that the squeezed probe provides better sensitivity across the whole detected spectrum. 

\begin{figure}
    \centering
    \includegraphics[width = 0.9\columnwidth]{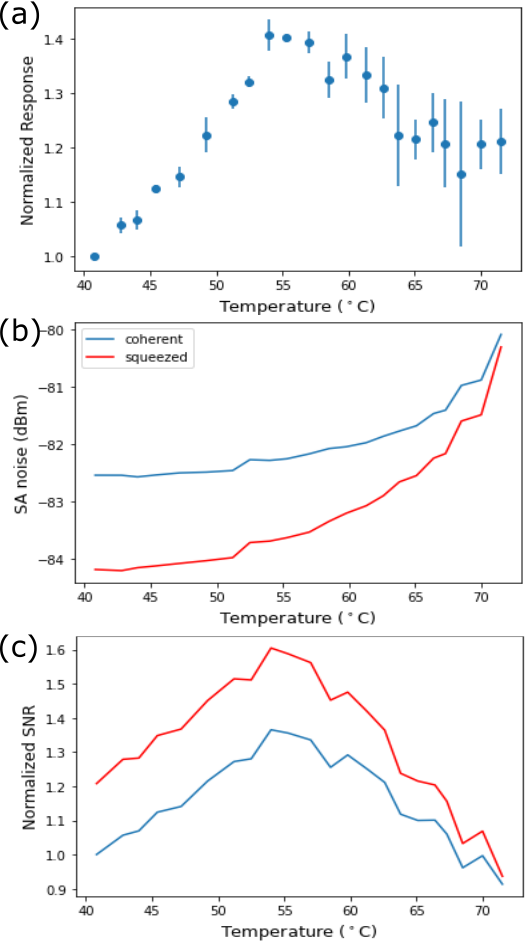}
    \caption{(a) Normalized magnetometer response as a function of the magnetometer cell temperature. The response is normalized to its value at $T=40.8^\circ$C. (b) Spectrum analyzer noise floor at the modulation frequency $\Omega_L=580$~kHz for the coherent and squeezed probe field as a function of the magnetometer cell temperature. Spectrum analyzer resolution and video bandwidth values are set to 1000~Hz and 1~Hz respectively. (c) Normalized quantum noise-limited estimated signal to noise ratio (SNR) as a function of the magnetometer cell temperature.  The SNR is normalized to its value for the coherent probe at 40.8$^\circ$C.}
    \label{fig:temperature noise}
\end{figure}

Lastly, we investigated how the magnetometer performance scales with the Rb vapor temperature. On one hand, the larger number of atoms participating in the interaction boosts the magneto-optical response. On the other hand, larger atomic density increases resonant losses, thus reducing the advantages of squeezing. For the measurements, we changed the temperature of the Rb cell from 40$^\circ$C, the minimum temperature to observe a significant signal, to 70$^\circ$C, where squeezing disappears. As expected, initially the signal grew with the temperature, reaching its maximum at 55$^\circ$C, and then slowly declined, as shown in Fig.~\ref{fig:temperature noise}(a). Since the noise measurements of the lock-in signal were very noisy, we instead recorded the noise floor from the spectrum analyzer signal for both squeezed and coherent probes. As shown in Fig.~\ref{fig:temperature noise}(b), this noise level increased with the atomic density for both coherent and squeezed probe fields at higher temperatures, likely due to the excess noise due to the PSR interaction generated independently in the magnetometer cell~\cite{NovikovaXiaoPRA2015}. Note, however, that the squeezed probe provided a lower noise level across the whole spectrum of temperatures. That observation confirms the finding in \cite{TroullinouPRL2021} that the modulated magnetometer scheme is not sensitive to the back-action noise. The increase of the noise level at lower temperatures is quite slow, while the signal scales much faster below 55$^\circ$C. As a result, if we estimate the temperature dependence of the signal to noise ratio (SNR) using the recorded spectrum analyzer noise level, we find that it largely follows the signal temperature dependence, peaking around  55$^\circ$C, as shown in Fig.~\ref{fig:temperature noise}(c). Since the resonant absorption did not noticeably below 55$^\circ$C as well, the deterioration of squeezing was minimal, causing noticeable improvement in estimated SNR for the squeezed probe field.


In conclusion, in this work we experimentally investigated the prospective of using a PSR squeezed quantum probe to improve the performance of an amplitude-modulation-based atomic magnetometer. While the demonstrated sensitivity of the magnetic field measurements In these experiments we were able to measure the value of a dc magnetic field of $0.8$~G with the estimated measurement uncertainty  $>250~\mathrm{pT}/\sqrt{\mathrm{Hz}}$. Such unimpressive sensitivity can be explained by strong power broadening and transitional broadening of the rotation resonance. On the other hand, we observed a clear noise reduction and the corresponding sensitivity improvement of 15\% when the polarization squeezed probe light was used for the detection. Our work can be particularly relevant to applications that require large operational bandwidth~\cite{Li2020} or cannot operate at high buffer gas pressure, such as a recently proposed quantum electron tracker~\cite{QET_D2detection_PhysRev2022}. 

\textit{Acknowledgements} We would like to thank Eugeniy Mikhailov for valuable advice and help, and Savannah Cuozzo for assistance with the squeezing probe arrangements. We acknowledge the support of the JLab LDRD grant for loaning us some essential equipment. 



\bibliography{references_magn.bib}





\end{document}